\documentclass[a4paper,11pt]{article}

\usepackage{pos}
\usepackage[]{units}
\usepackage{multirow}
\usepackage{lipsum} %package to generate placeholder text in the following
\usepackage{lineno}
\title{Electric-Field Reconstruction for Radio Detection of Inclined Air Showers in Three Polarizations}

\ShortTitle{Electric-Field Reconstruction in Three Polarizations}

\author*[a,b]{Kewen Zhang}
\author[c]{Lukas Gülzow}
\author[c,d]{Tim Huege}
\author[e,f,l]{Ramesh Koirala}
\author[a,b]{Pengxiong Ma}
\author[g,h]{Matías Tueros}
\author[i]{Xin Xu}
\author[e,f,j,k]{Chao Zhang}
\author[i]{Pengfei Zhang}
\author[a,b] {Yi Zhang}
\onbehalf{for the GRAND Collaboration{\normalsize \\ \normalfont(a complete list of authors can be found at the end of the proceedings)}\\}

\affiliation[a]{Key Laboratory of Dark Matter and Space Astronomy, Purple Mountain Observatory}
\affiliation[b]{School of Astronomy and Space Science, University of Science and Technology of China}
\affiliation[c]{Institute for Astroparticle Physics (IAP), Karlsruhe Institute of Technology}
\affiliation[d] {Astrophysical Institute, Vrije Universiteit Brussels}
\affiliation[e]{School of Astronomy and Space Science, Nanjing University}
\affiliation[f]{Key laboratory of Modern Astronomy and Astrophysics, Nanjing University}
\affiliation[g]{IFLP - CCT La Plata - CONICET}
\affiliation[h]{Depto. de Física, Fac. de Cs. Ex., Universidad Nacional de La Plata}
\affiliation[i]{School of Electronic Engineering, Xidian University}
\affiliation[j] {Department of Astronomy, School of Physics, Peking University}
\affiliation[k]{Kavli institute for astronomy and astrophysics, Peking University}
\affiliation[l]{Space Research Centre, Faculty of Technology, Nepal Academy of Science and Technology (NAST)}
\emailAdd{kwzhang@pmo.ac.cn}

\abstract{

Accurate reconstruction of the electric field produced by extensive air showers is essential for the radio-detection technique, as the key parameters of interest of the primary particles that generated the showers are the amplitude, polarization, frequency spectrum, and energy fluence carried by the electric field at each  receiving radio antenna. Conventional electric-field reconstruction methods primarily focus on antennas with two horizontal polarizations. In this work, we introduce an analytic  $\chi^2$ minimization method that is applicable to both two and three polarizations. This solution has been verified for simple and realistic antenna responses, with a particular focus on inclined air showers. Our method achieves standard deviations better than 4\% and 6\% for the estimation of the Hilbert peak envelope amplitude of the electric field and the energy fluence, respectively, with an antenna-response-dependent bias. Additionally, we have studied the dependence of the method with arrival direction showing that it has a good performance in the zenith range from 63$^\circ$ up to 80$^\circ$. This work also demonstrates that incorporating vertically polarized antennas enhances the precision of the reconstruction, leading to a more accurate and reliable electric-field estimation for inclined air showers.

\vspace{4mm}

}

\ConferenceLogo{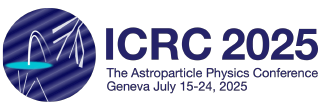}

\FullConference{39th International Cosmic Ray Conference (ICRC2025)\\  15--24 July, 2025\\ Geneva, Switzerland}

\begin{document}

\maketitle

\section{Introduction}
\label{intro}

The reconstruction of extensive-air-shower (EAS) properties from radio measurements fundamentally depends on an accurate knowledge of the incident electric field. Since antennas record voltage signals rather than the electric field directly, a crucial first step in reliable radio-based analyses is to reconstruct the electric field from these voltage traces. The most widely-used technique for reconstructing the electric field from voltage traces, involving the inversion of a 2×2 response matrix, was developed by AERA \cite{Abreu_2011}. This technique was further improved in \cite{Glaser_2019, welling2019reconstructing}, where a forward-folding technique using multiple channel measurements was developed to improve the reconstruction accuracy for signals with low signal-to-noise ratio (SNR). A more recent technique employs information field theory to reconstruct low-SNR signals with high precision \cite{welling2021reconstructing,simon2024information}. These methods, however, have been applied primarily to antennas with two horizontal polarizations, leading to an incomplete sampling of the three-dimensional electric field. For the case of three polarizations, the LOPES-3D experiment pioneered multiple weighted-reconstruction methodologies, including polarization-direction-specific weighting factors calibrated according to their sensitivity to incident angles \cite{Huber2014_1000043289, huber2022lopes}, with no significant improvement relative to the conventional matrix inversion method.

In this context, we introduce a novel approach based on an analytic $\chi^2$ minimization for reconstructing the electric field, inspired by \cite{welling2019reconstructing}. It incorporates not only the two horizontal polarizations but also the additional vertical polarization, allowing a more complete and precise reconstruction of the electric field without assumptions on the signal shape, and thus yielding a robust electric field estimation even under varying signal conditions and antenna responses.

\section{Simulations}
\label{sims}
This work is based on detailed EAS simulations, including electric-field generation, antenna response, and galactic noise. By combining these components, we produced realistic voltage signals for robust evaluation of electric-field reconstruction.

\subsection{Simulation of cosmic-ray radio signals}
We used ZHAireS \cite{Alvarez_Mu_iz_2012} to simulate radio signals from EAS induced by proton (50\%) and iron (50\%) primaries. The dataset includes 4,160 EAS events which spans zenith angles from 63.0$^\circ$ to 87.1$^\circ$ in uniform $\log_{10}(1/\cos \theta) = 0.08$ steps, azimuths from 0$^\circ$ to 180$^\circ$ in 45$^\circ$ steps, and energies from 0.126 to 3.98 EeV in $\log_{10}(E/\mathrm{EeV}) = 0.1$ steps. Simulations were performed for a radio-quiet site near Dunhuang, China, using a 56 $\mu$T geomagnetic field (61$^\circ$ inclination) and the extended Linsley US-standard atmosphere \cite{Heck:1998vt} with an exponential refractive index profile (scale height 8.2 km, sea-level index 1.000325).

 A star-shaped antenna array, centered on the shower core and defined in the shower plane, was projected onto the ground, resulting in 160 antennas arranged along eight arms with 20 antennas each. The inter-antenna spacing of the arms was optimized for the Cherenkov angle to ensure maximum signal coverage. For each antenna, electric-field components were recorded in three polarization directions: $E_x$ (South-North), $E_y$ (East-West), and $E_z$ (vertical). Signals were sampled with a 0.5 ns time bin over a 1000 ns window, yielding a frequency resolution of 1 MHz after Fourier transformation.

\subsection{Antenna responses}
\label{ant_resp}

We evaluate two antenna designs for three-polarization electric-field reconstruction: a simple three-arm dipole and the HORIZON antenna \cite{GRAND:2018iaj, GRAND:2023mco}. The dipole serves as a baseline, while HORIZON represents a realistic detector setup. Both are placed 3 m above ground, with responses simulated using Ansys HFSS \cite{ansysHFSS}, including ground effects.

Each antenna has three orthogonal arms (South-North, East-West, vertical). The dipole uses 2.6 m arms and operates in the 30–80 MHz band. The HORIZON antenna features a butterfly-shaped steel radiator optimized for inclined EAS, operating in the 50–200 MHz band with an impedance-matching network.

%The dipole antenna consists of three orthogonal arms aligned along the South-North, East-West, and vertical directions. Each arm is 2.6 meters long (two 1.3-meter oscillators) and operates in the 30–80 MHz band. The HORIZON antenna, on the other hand, features a symmetric butterfly-shaped steel radiator with three orthogonal arms, optimized for detecting inclined showers. Operating between 50–200 MHz, it includes two radiators in the horizontal arms and one in the vertical, with an impedance-matching network.
%, similar to the design used in the LOPES-3D experiment \cite{Huber2014_1000043289, huber2022lopes}.
% This design prioritizes phase-center stability over omnidirectionality and is based on earlier CODALEMA prototypes \cite{CHARRIER2012S142}, refined for the GRAND experiment \cite{GRAND:2018iaj, GRAND:2023mco}.

\subsection{Background noise}
\label{bkg_noise}

Accurate electric-field reconstruction is challenged by background noise. While remote sites reduce human-made radio-frequency interference (RFI), residual contamination from satellites, aircraft, and FM radio remains. These narrowband sources are mitigated via filtering and RFI suppression. Transient phenomena like thunderstorms and solar flares are excluded from this study due to their episodic nature. In the considered frequency range, the dominant irreducible noise is galactic synchrotron radiation \cite{alvarez2020giant, busken2023uncertainties}.

The background spectrum across 30–200 MHz is modeled using LFmap \cite{polisensky2007lfmap}, incorporating both galactic and extragalactic emissions. The resulting noise level varies with local sidereal time (LST); in this study, we adopt the noise corresponding to LST = 18 h, when it reaches a relatively high value (approximately $20\ \mu\mathrm{V}$) compared to other times. %The sky temperature follows a power law, $T_\mathrm{sky}(f)\propto f^{-\beta}$, with $\beta$ \cite{Guzm_n_2010}. The spectral brightness is converted via the Rayleigh-Jeans law: $B_\mathrm{f}=\frac{2k_\mathrm{B} f^2 T_\mathrm{sky}}{c^2}$, where $k_B$ is the Boltzmann constant, and $c$ is the speed of light. and the sky noise power is given by: $P_\mathrm{sky}(t',f)=\frac{1}{2}\int B_f(\theta,\varphi,t') A_e(\theta,\varphi,f)\sin{\theta}\mathrm{d}\theta \mathrm{d}\varphi $, where $t'$ represents the the local sidereal time (LST), where $A_e = \frac{\lambda^2}{4\pi}G$ is the antenna’s effective aperture and G is the gain with ideal antenna efficiency $\eta_A$ = 1. Assuming an impedance-matched condition with $Z_0  = 50 \Omega$, the voltage induced by Galactic noise becomes: $ V_\mathrm{sky}=2\sqrt{Z_0 P_\mathrm{sky}}$. 
To simulate realistic time-domain traces, uniformly distributed random phases were assigned in the frequency domain before applying an inverse fast Fourier transform (iFFT) to the modeled spectra. We assume that the input to the reconstruction algorithm is the open-circuit voltage $V_\mathrm{oc}$ at the antenna output. Electronic noise between $V_\mathrm{oc}$ and an amplified, digitized signal $V_\mathrm{ADC}$ is not modeled here but will be addressed in future work with experimental data.

\section{Electric-field reconstruction}
This study employs two electric-field reconstruction methods. The first method is a conventional matrix-inversion technique, commonly used in radio detection experiments \cite{Abreu_2011}, which inverts the antenna response matrix using two or three polarizations (Section \ref{conv_mthd}). The second method is a novel analytic $\chi^2$ minimization method utilizing all three polarizations, introduced for the first time in this work (Section \ref{least_estm}).

\subsection{Matrix-inversion method}
\label{conv_mthd}
\begin{figure}[htbp]
\centering
\includegraphics[trim={0.5cm 3cm 0.5cm 3cm}, width=0.4\textwidth]{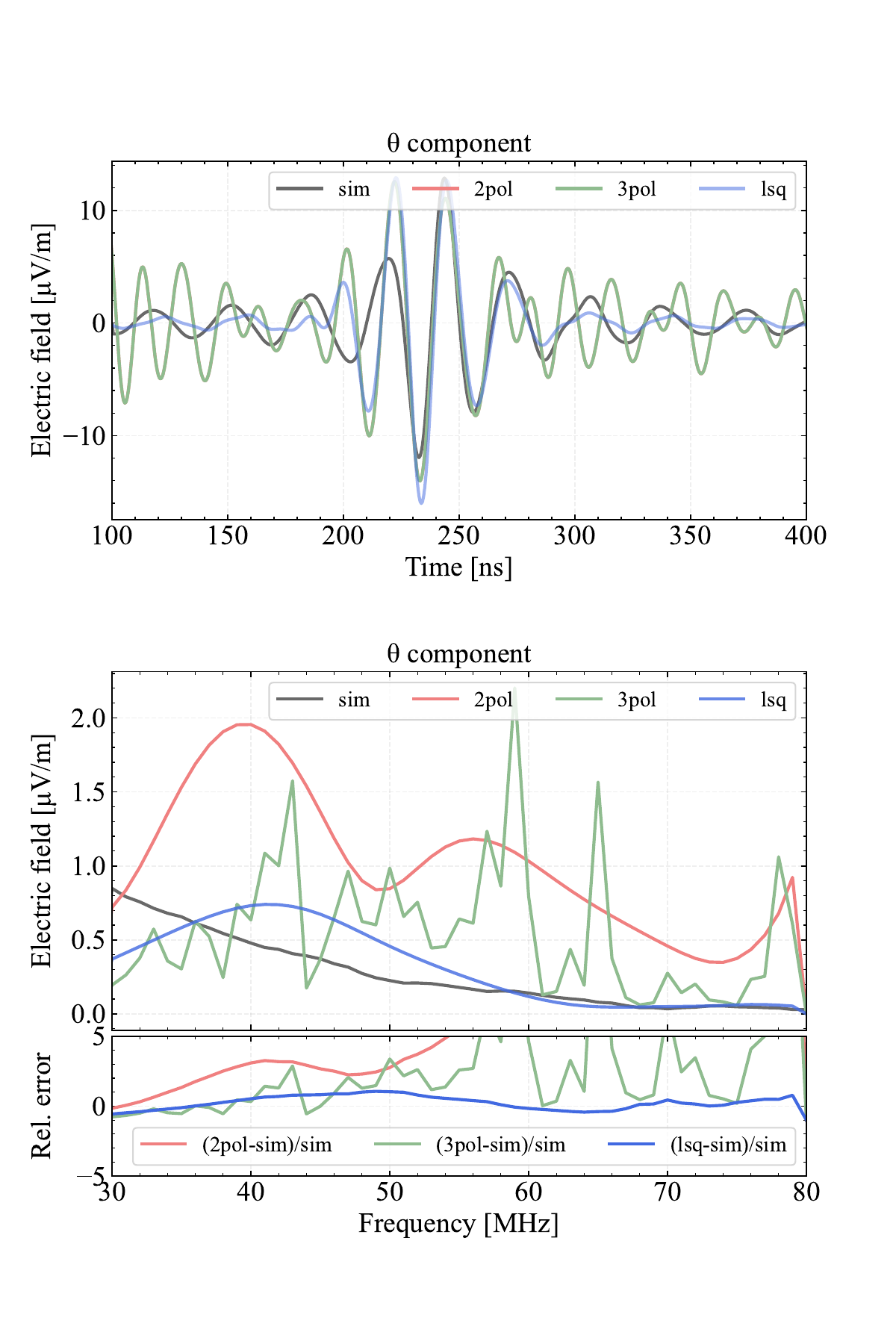}
\includegraphics[trim={0.5cm 3cm 0.5cm 3cm}, width=0.4\textwidth]{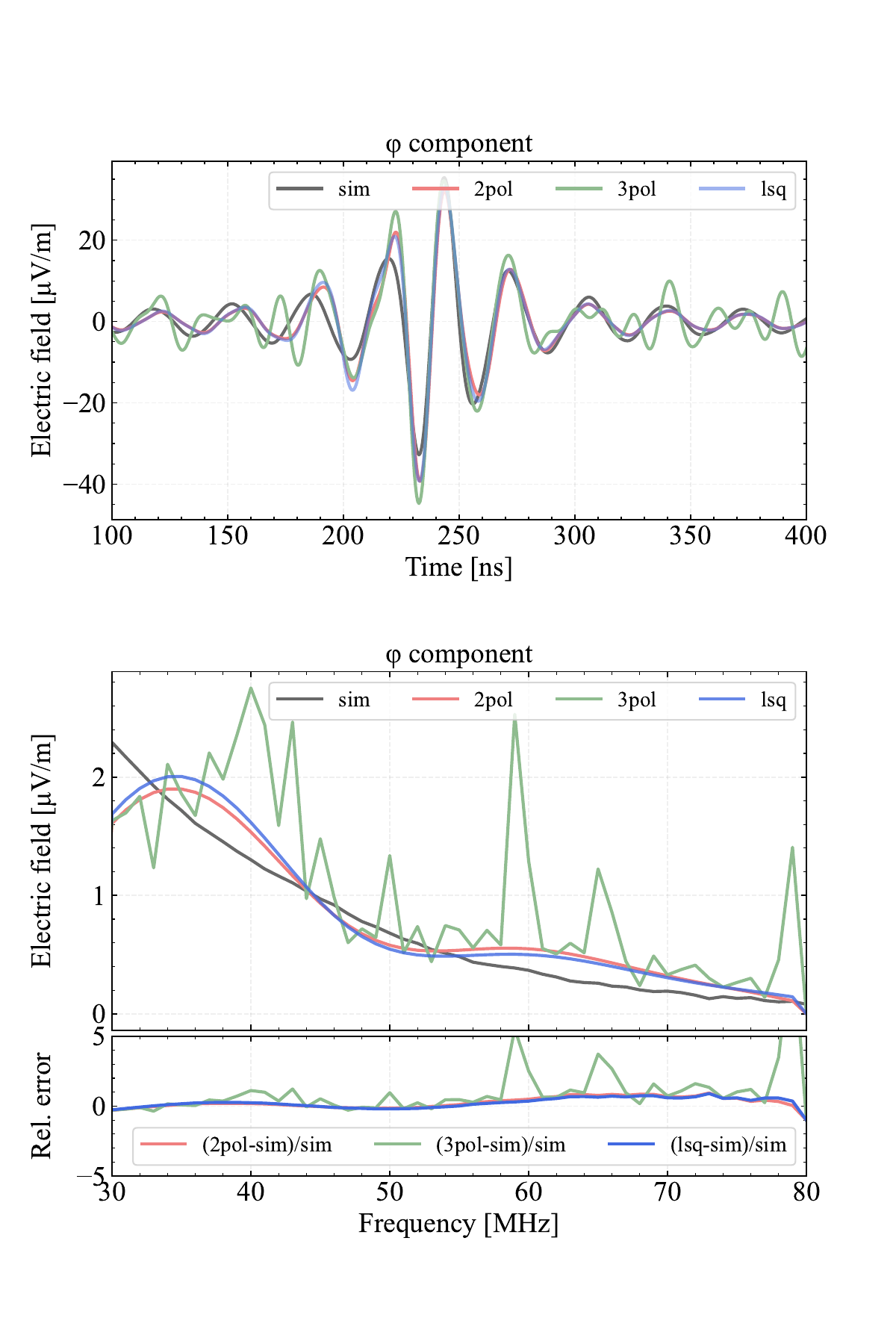}    
\caption{Comparison of the matrix-inversion method, for two (2pol) and three (3pol) polarizations, and the analytic $\chi^2$ minization method (lsq), for the simple dipole antenna. \textit{Top}: Time traces of the electric field. \textit{Bottom}: Electric-field spectrum in the frequency domain.}
    \label{fig:e_rec_dipole}
\end{figure}
To assess the performance of the conventional matrix-inversion method for the simple dipole antenna, we use an example voltage trace with an SNR of approximately 6 in each polarization. Figure 1 shows the corresponding simulated electric field, as well as the reconstructed electric field using both two and three polarizations The normalized cross-correlation values \cite{corstanje2023high}, ranging between -1 (perfectly anticorrelated) and +1 (perfectly correlated),  between the reconstructed and simulated electric field traces are used to assess the reconstruction performance. Although the correlation of the $\theta$-component improves from -0.13 to 0.40 when including the third polarization, the reconstructed frequency spectrum remains poor, as shown in the bottom left panel of Figure \ref{fig:e_rec_dipole}. The correlation of the $\varphi$-component decreases from 0.93 to 0.80 when including the third polarization, and also introduces artifacts in the frequency spectrum (bottom right panel of Figure \ref{fig:e_rec_dipole}). These results suggest that including the third polarization benefits the weaker $\theta$-component but slightly degrades the stronger $\varphi$-component due to artifacts introduced by the small $r$-component. This highlights the need for further refinement to achieve a stable and accurate reconstruction.

\subsection{Analytic $\chi^2$ minimization method}
\label{least_estm}
In this study, we build a single $\chi ^2$ for each antenna with three polarizations, without incorporating any parameters related to the signal characteristics.  Consequently, it offers a more direct and model-independent framework compared to alternative approaches, enhancing the simplicity of the reconstruction. Given the measured voltage, noise spectrum, and antenna response, the \(\chi^2\) is formulated in frequency domain as

\begin{equation}
\begin{aligned}
    \chi^2 & = \sum_{i=1}^3 \left( \frac{{\mathcal{V}}_i - \mathcal{H}_i \begin{pmatrix} \mathcal{E}_\theta \\ \mathcal{E}_\varphi \end{pmatrix} }{\sigma_{\mathcal{V}_i}} \right)^2  = (\boldsymbol{\mathcal{V}} - \boldsymbol{\mathcal{H}} \boldsymbol{\mathcal{E}})^T \sigma_{\mathcal{V}}^{-1} (\boldsymbol{\mathcal{V}} - \boldsymbol{\mathcal{H}} \boldsymbol{\mathcal{E}}),
\end{aligned}
\end{equation}

where $\mathcal{V}$ denotes the open-circuit voltage, $\mathcal{H}$ represents the antenna response, and $\sigma_{\mathcal{V}}$ is the background noise level. The subscript $i$ indicates the polarization channel, with three components in total for our setup. The electric-field vector is denoted by $\boldsymbol{\mathcal{E}}$, with subscripts $\theta$ and $\varphi$ referring to its zenith and azimuth components, respectively. $\mathcal{E}_r$ is neglected here as it corresponds to the propagation direction of the transverse wave.

 We assume that the error in this measurement is attributed to background noise. Consequently, the covariance matrix is constructed as a diagonal matrix $\sigma_\mathcal{V}=\operatorname{diag}\left(\sigma_{\mathcal{V} 1}, \sigma_{\mathcal{V} 2}, \sigma_{\mathcal{V} 3}\right)$, where the diagonal elements correspond to the squared noise spectrum in each polarization of the antenna. 

To minimize $\chi^2$, we compute $\nabla_{\mathcal{E}} \chi^2 = -2 (\boldsymbol{\mathcal{H}}^T \sigma_{\mathcal{V}}^{-1} \boldsymbol{\mathcal{V}} - \boldsymbol{\mathcal{H}}^T \sigma_{\mathcal{V}}^{-1} \boldsymbol{\mathcal{H}}\boldsymbol{\mathcal{E}}) = 0$. From this, we obtain an analytic  solution of the electric field, given by $\boldsymbol{\mathcal{E}} = (\boldsymbol{\mathcal{H}}^T \sigma_{\mathcal{V}}^{-1} \boldsymbol{\mathcal{H}})^{-1} \boldsymbol{\mathcal{H}}^T \sigma_{\mathcal{V}}^{-1} \boldsymbol{\mathcal{V}}$.

\subsection{Matrix-inversion vs. analytic $\chi^2$ minimization}
\label{comparison1}

To assess reconstruction quality, we  first compare the analytic $\chi^2$ minimization method with the matrix-inversion method as shown in Figure \ref{fig:e_rec_dipole}. For the dipole antenna, the matrix-inversion approach with three polarizations yields cross-correlation coefficients of 0.40 for the $\theta$-component and 0.80 for the $\varphi$-component. In contrast, the $\chi^2$ minimization method achieves significantly higher cross correlations--0.77 for the $\theta$-component and 0.95 for the $\varphi$-component--demonstrating its improved reconstruction accuracy.  %As shown in Figure \ref{fig:e_rec_dipole_lsq}, the $\chi^2$ method also effectively suppresses spikes in the reconstructed signals.

%\subsection{Application to the HORIZON antenna}

\begin{figure}[htbp]
\centering
\includegraphics[trim={0.5cm 3cm 0.5cm 3cm}, width=0.4\textwidth]{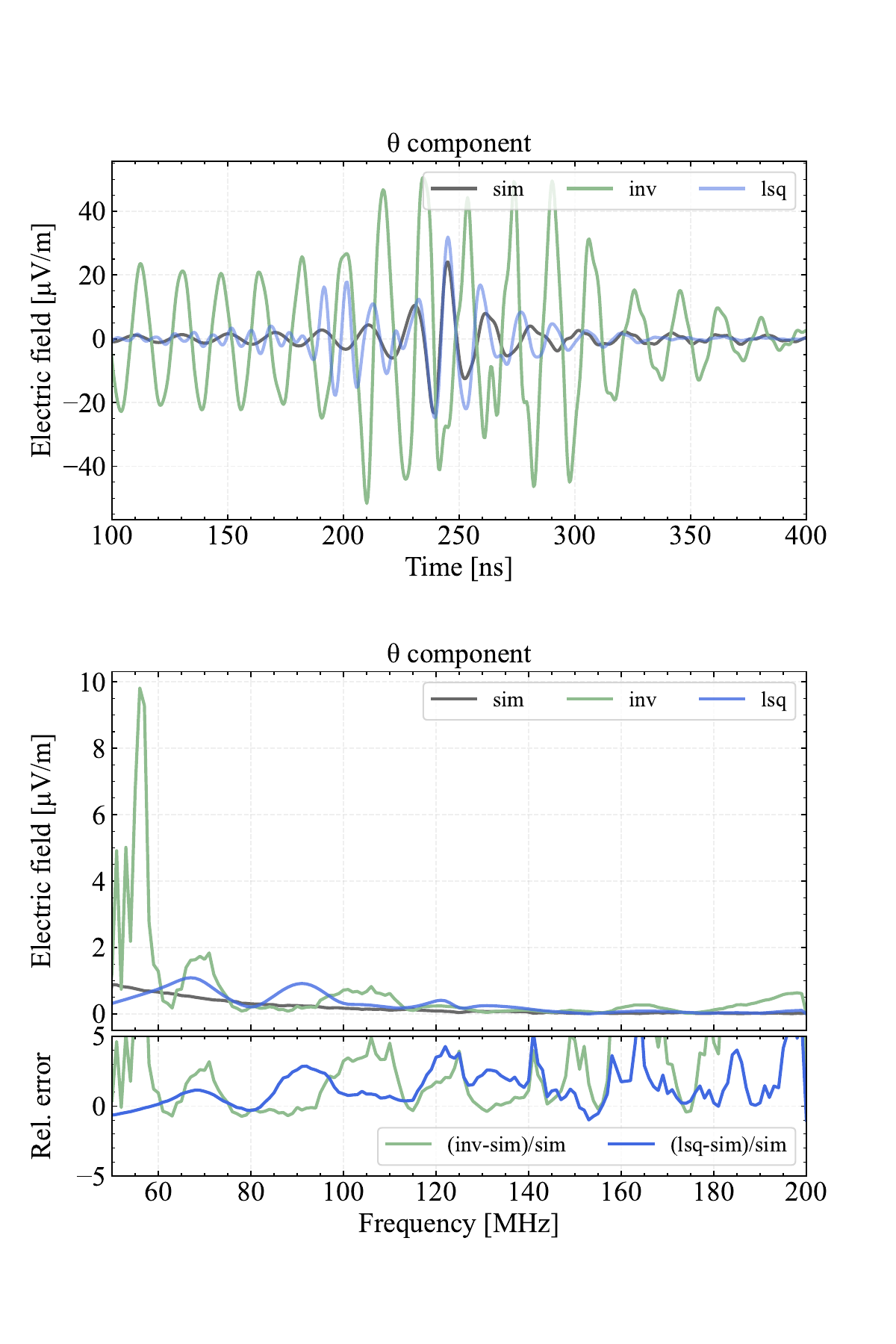}
\includegraphics[trim={0.5cm 3cm 0.5cm 3cm}, width=0.4\textwidth]{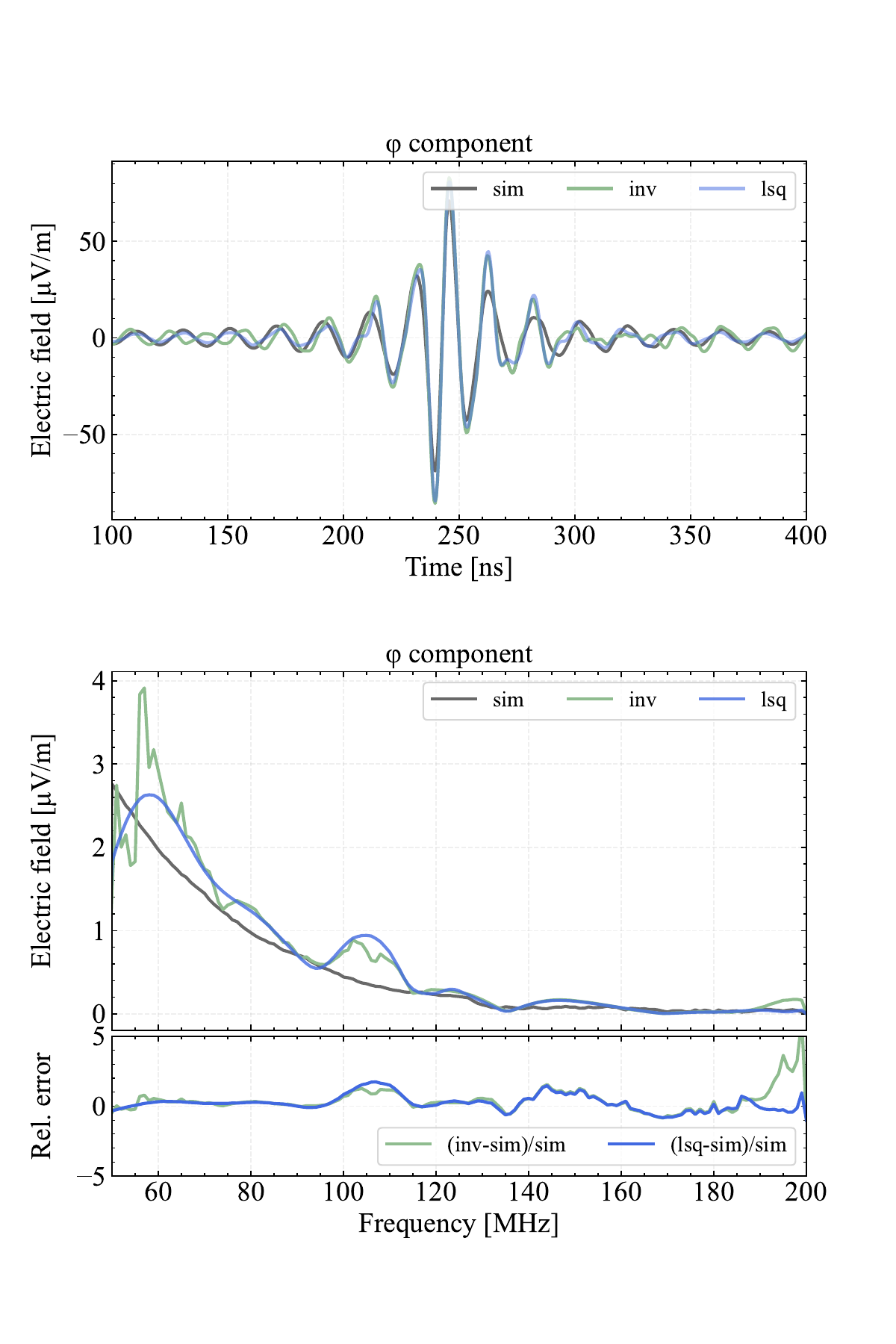}    
    \caption{Comparison of the matrix inversion method (inv) and the analytic $\chi^2$ minimization (lsq) reconstruction methods in three polarizations applied to our example electric-field trace (sim), for the HORIZON antenna. \textit{Top}: time traces of the electric field. \textit{Bottom}: electric-field spectrum in the frequency domain. }
    \label{fig:e_rec}
\end{figure}

We further apply the analytic $\chi^2$ minimization method to the HORIZON antenna to test thet method's robustness under varying antenna responses. For this test, we use the same EAS simulation as used for Figure \ref{fig:e_rec_dipole}, but we take our test trace at a different antenna position to ensure an SNR $\sim$ 6 at the voltage level. The simulated and reconstructed electric fields are shown in Figure \ref{fig:e_rec}. The matrix inversion method gives cross-correlation coefficients of –0.34 (for $\theta$) and 0.96 (for $\varphi$), while the $\chi^2$ minimization method achieves a significantly better performance, with cross-correlation values of 0.79 and 0.97, respectively. 
These results confirm the method’s improved and consistent performance across different antenna designs. In both the frequency and time domains, the reconstruction based on $\chi^2$ minimization aligns more closely with the simulated signal.

\section{Statistical performance of the analytic  $\chi^2$ minimization method}

We evaluate the resolution of the analytic $\chi^2$ minimization method using the simulation libary described in Section \ref{sims}. Only events meeting the following criteria in a minimum of five antennas are selected: an SNR greater than 5 in at least one antenna arm, a peak time ($t_{\text{peak}}$) within \unit[$\pm$200]{ns} of the expected signal time, and the use of only the innermost 16 antennas per arm to exclude weak signals.
\subsection{Comparison of the peak envelope amplitudes}
\label{hilbert_cmp}
\begin{figure}[!htbp]
\centering
\includegraphics[width=1.00\textwidth]{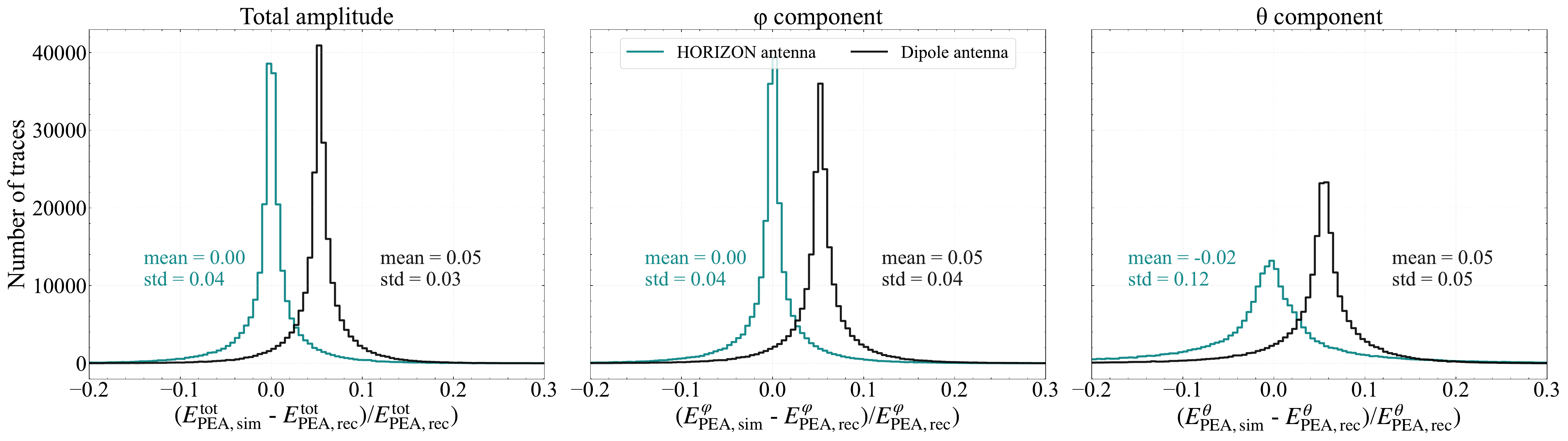}
\caption{Relative error of reconstructed (rec) PEA w.r.t. simulated (sim) PEA for total (left), $\varphi$ (middle), and $\theta$ (right) electric-field components. Results are shown for the HORIZON (green) and dipole (black) antennas, with statistical summaries in matching colors.}
\label{fig:e_comp}
\end{figure}

The Hilbert envelope provides a smooth representation of a signal, making it useful for analyzing noisy, oscillatory data. To assess the reconstruction performance, we compare the peak envelope amplitude (PEA) of the simulated and reconstructed electric-fields, for both the dipole and HORIZON antenna models. The relative error distributions are shown in Figure~\ref{fig:e_comp}.

%For both antennas, we obtain an accurate PEA reconstruction, with relative-error distributions centered near zero. The $\varphi$-component is consistently better reconstructed than the $\theta$-component, due to stronger signals and higher antenna gain across most frequencies. This difference is more evident for HORIZON antenna.

%For the dipole, the relative error has a standrad deviation of 0.04 and a small bias (mean/median of 0.05), indicating a slight underestimation of the true value. For the HORIZON antenna, the bias is negligible (mean/median near 0), but the error spread is marginally larger, especially in the $\theta$-component (std = 0.12), mainly due to a few outlier traces from lower-SNR (SNR < 10) events.

For both antenna types, we find that the PAE reconstruction of the analytic $\chi^2$ minimization method is precise and robust. The standard deviation of the relative-error distributions is generally better than 5\%, with the exception of the $\theta$-component reconstruction of the HORIZON antenna, which yields a standard deviation of 12\%. This is due to the weaker $\theta$-component of the simulated electric fields, as well as a lower antenna gain along theta across most frequencies, which is more prominent for the HORIZON antenna.

We also find that the relative-error distributions of the HORIZON antenna are centered around 0, while those of the dipole antenna display a consistent $\sim$5\% bias, indicating a slight underestimation of the the simulated electric-field values. This bias may result from several factors—such as antenna response, frequency coverage, and directional sensitivity—which are still under investigation.

Overall, the analytic $\chi^2$ minimization method demonstrates robust reconstruction performance for both antenna types.

\subsection{Comparison of energy fluence}
The electromagnetic component of an air shower dominates the primary energy and can be inferred from the radiation energy \cite{Glaser:2016qso}. At each antenna, the reconstructed electric field is used to calculate the energy fluence—i.e., the energy deposited per unit area—via $\Phi = c \cdot \epsilon_0 \cdot  \int \vec{E}^2(t) dt$, with $\vec{E}$(t) the electric field, $c$ the speed of light, and $\epsilon_0$ the vacuum permittivity. The total radiation energy is obtained by integrating the energy fluence across the array.

To isolate signal from noise, we integrate over a 100 ns window centered at the pulse peak and subtract the fluence from a separate 100 ns window at the trace end. We show the relative error of the square root of the fluence, $\sqrt{\Phi}$, as it is proportional to the shower energy. The resulting energy fluence shows similar trends to the peak envelope amplitude (PEA), but with amplified discrepancies due to error propagation in squaring and integrating the reconstructed electric-field. As such, we obtain typical standard deviations on the relative error distributions better than 6\%, with the exception of the $\theta$-component of the fluence, which has a relative error of 19\%.

\begin{figure}[htbp]
\hspace{2em}
\centering
\includegraphics[width=1.03\textwidth]{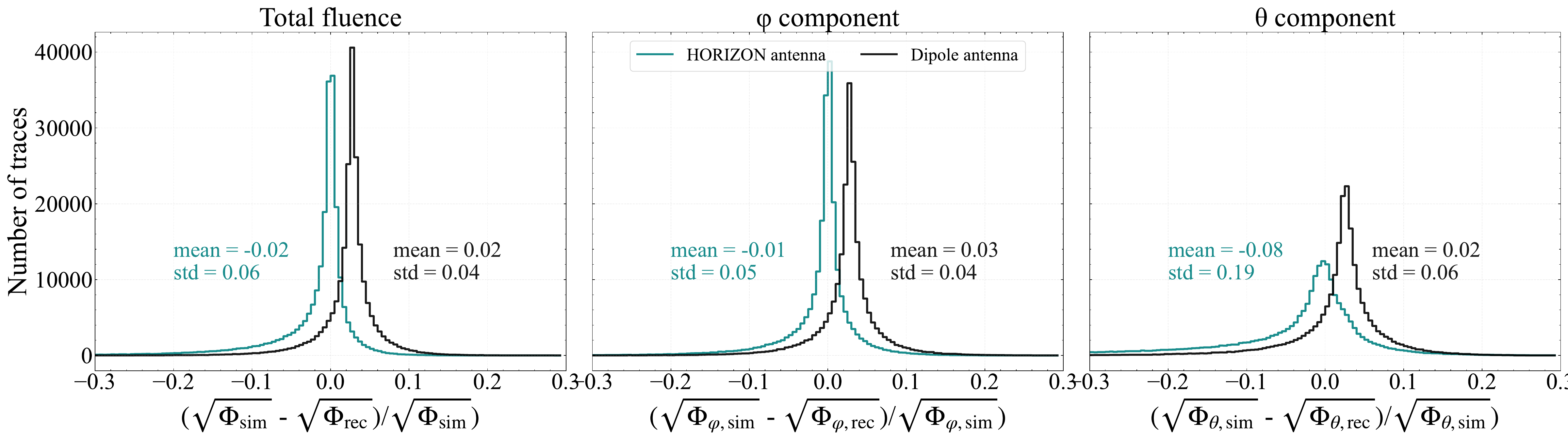}
\caption{Relative error of the fluence ($\Phi$) reconstruction, comparing the reconstructed $\sqrt{\Phi_{\mathrm{rec}}}$ with the simulated $\sqrt{\Phi_{\mathrm{sim}}}$ for total (left), $\varphi$ (middle), and $\theta$ (right) electric-field components. Results are shown for the HORIZON (green) and dipole (black) antennas, with statistical summaries in matching colors.}
\label{fig:fluence_cmp}
\end{figure}

Overall, both antennas achieve good fluence reconstruction precision using the analytic $\chi^2$ method, especially for SNR$>$10. The dipole shows a consistent 3\% underestimation; HORIZON has near-zero bias. The $\theta$-component shows larger variance due to its weaker signal strength. These results confirm the robustness of the method, although SNR-dependent biases, especially at low SNR, require further calibration for real data.

\section{Conclusions and outlook}

We developed an analytic $\chi^2$ minimization method to improve the reconstruction of the electric field from air-shower radio signals. This approach avoids assumptions about signal properties and provides robust performance across both dipole and HORIZON antenna models with three polarizations. Compared to the conventional matrix-inversion technique, our method better handles uncertainties at low-gain or low-SNR frequencies, reducing the typical standard deviation to 4\% for PEA and $\lesssim$6\% for energy fluence.

Our method was tested with ZHAireS simulations, including simulated galactic noise. Future validation using experimental noise is planned. Additionally, while we used the 30–200 MHz band—common in current experiments—investigating the method's performance across sub-bands or broader frequency ranges remains a direction for future work. Furthermore, an iterative reconstruction involving updated direction inputs could further refinehe accuracy of our method.

% Bibtex references:
%\bibliographystyle{ICRC}
%\bibliography{references}

\providecommand{\href}[2]{#2}\begingroup\raggedright\endgroup

% Alternatively, you can include references by hand:
%\begin{thebibliography}{99}
%\bibitem{...}
%
%\end{thebibliography}

\clearpage

%The following list of authors, affiliations and funding agencies will be updated at the day of submission. The following template is a placeholder generated with the member list updated to April 1, 2025..
\section*{Full Author List: GRAND Collaboration}

\scriptsize
\noindent
J.~Álvarez-Muñiz$^{1}$, R.~Alves Batista$^{2, 3}$, A.~Benoit-Lévy$^{4}$, T.~Bister$^{5, 6}$, M.~Bohacova$^{7}$, M.~Bustamante$^{8}$, W.~Carvalho$^{9}$, Y.~Chen$^{10, 11}$, L.~Cheng$^{12}$, S.~Chiche$^{13}$
% ...
\\
\\
$^{1}$Departamento de Física de Particulas \& Instituto Galego de Física de Altas Enerxías, Universidad de Santiago de Compostela, 15782 Santiago de Compostela, Spain \\
$^{2}$Institut d'Astrophysique de Paris, CNRS  UMR 7095, Sorbonne Université, 98 bis bd Arago 75014, Paris, France \\
$^{3}$Sorbonne Université, Université Paris Diderot, Sorbonne Paris Cité, CNRS, Laboratoire de Physique 5 Nucléaire et de Hautes Energies (LPNHE), 6 4 place Jussieu, F-75252, Paris Cedex 5, France \\
$^{4}$Université Paris-Saclay, CEA, List,  F-91120 Palaiseau, France \\
$^{5}$Institute for Mathematics, Astrophysics and Particle Physics, Radboud Universiteit, Nijmegen, the Netherlands \\
$^{6}$Nikhef, National Institute for Subatomic Physics, Amsterdam, the Netherlands \\
$^{7}$Institute of Physics of the Czech Academy of Sciences, Na Slovance 1999/2, 182 00 Prague 8, Czechia \\
$^{8}$Niels Bohr International Academy, Niels Bohr Institute, University of Copenhagen, 2100 Copenhagen, Denmark \\
$^{9}$Faculty of Physics, University of Warsaw, Pasteura 5, 02-093 Warsaw, Poland \\
$^{10}$Key Laboratory of Dark Matter and Space Astronomy, Purple Mountain Observatory, Chinese Academy of Sciences, 210023 Nanjing, Jiangsu, China \\
$^{11}$School of Astronomy and Space Science, University of Science and Technology of China, 230026 Hefei Anhui, China \\
$^{12}$National Astronomical Observatories, Chinese Academy of Sciences, Beijing 100101, China \\
$^{13}$Inter-University Institute For High Energies (IIHE), Université libre de Bruxelles (ULB), Boulevard du Triomphe 2, 1050 Brussels, Belgium \\
% ...

%%%%%%%%%%%%%%%%%%%%%%%%%%%%%%%%%%%%%%%%%%%%%%%%%%%%%%%%%%%%%%
%%%%%%%%%%%%%%%%%%%%%%%%%%%%%%%%%%%%%%%%%%%%%%%%%%%%%%%%%%%%%%

\subsection*{Acknowledgments}

\noindent
The GRAND Collaboration is grateful to the local government of Dunhuag during site survey and deployment approval, to Tang Yu for his help on-site at the GRANDProto300 site, and to the Pierre Auger Collaboration, in particular, to the staff in Malarg\"ue, for the warm welcome and continuing support.
The GRAND Collaboration acknowledges the support from the following funding agencies and grants.
%%%%
\textbf{Brazil}: Conselho Nacional de Desenvolvimento Cienti\'ifico e Tecnol\'ogico (CNPq); Funda\c{c}ão de Amparo \`a Pesquisa do Estado de Rio de Janeiro (FAPERJ); Coordena\c{c}ão Aperfei\c{c}oamento de Pessoal de N\'ivel Superior (CAPES).
%%%%
\textbf{China}: National Natural Science Foundation (grant no.~12273114); NAOC, National SKA Program of China (grant no.~2020SKA0110200); Project for Young Scientists in Basic Research of Chinese Academy of Sciences (no.~YSBR-061); Program for Innovative Talents and Entrepreneurs in Jiangsu, and High-end Foreign Expert Introduction Program in China (no.~G2023061006L); China Scholarship Council (no.~202306010363); and special funding from Purple Mountain Observatory.
%%%%
\textbf{Denmark}: Villum Fonden (project no.~29388).
%%%%
\textbf{France}: ``Emergences'' Programme of Sorbonne Universit\'e; France-China Particle Physics Laboratory; Programme National des Hautes Energies of INSU; for IAP---Agence Nationale de la Recherche (``APACHE'' ANR-16-CE31-0001, ``NUTRIG'' ANR-21-CE31-0025, ANR-23-CPJ1-0103-01), CNRS Programme IEA Argentine (``ASTRONU'', 303475), CNRS Programme Blanc MITI (``GRAND'' 2023.1 268448), CNRS Programme AMORCE (``GRAND'' 258540); Fulbright-France Programme; IAP+LPNHE---Programme National des Hautes Energies of CNRS/INSU with INP and IN2P3, co-funded by CEA and CNES; IAP+LPNHE+KIT---NuTRIG project, Agence Nationale de la Recherche (ANR-21-CE31-0025); IAP+VUB: PHC TOURNESOL programme 48705Z. 
%%%%
\textbf{Germany}: NuTRIG project, Deutsche Forschungsgemeinschaft (DFG, Projektnummer 490843803); Helmholtz—OCPC Postdoc-Program.
%%%%
\textbf{Poland}: Polish National Agency for Academic Exchange within Polish Returns Program no.~PPN/PPO/2020/1/00024/U/00001,174; National Science Centre Poland for NCN OPUS grant no.~2022/45/B/ST2/0288.
%%%%
\textbf{USA}: U.S. National Science Foundation under Grant No.~2418730.
%%%
Computer simulations were performed using computing resources at the CCIN2P3 Computing Centre (Lyon/Villeurbanne, France), partnership between CNRS/IN2P3 and CEA/DSM/Irfu, and computing resources supported by the Chinese Academy of Sciences.

\end{document}